\documentclass[fleqn,usenatbib]{mnras}

\usepackage[T1]{fontenc}
\usepackage{ae,aecompl}
\usepackage{xcolor}
\usepackage{comment}

\usepackage{graphicx}	
\usepackage{amsmath}	
\usepackage{amssymb}	
\usepackage{pdflscape}	
\usepackage[english]{babel}
\pdfoutput=1

\newcommand{\unitnormal}{\hat{\mathbf{n}}}

\newcommand{\kms}{\rm{km\,s^{-1}}}
\newcommand{\cms}{\rm{cm\,s^{-1}}}
\newcommand{\rsol}{R_{\rm \odot}}

\title[Photogravimagnetic assists of light sails]{Photogravimagnetic assists of light sails: a mixed blessing for Breakthrough Starshot?}

\author[D. H. Forgan et al. ]
{Duncan H. Forgan$^{1,2}$\thanks{Contact e-mail: \href{mailto:dhf3@st-andrews.ac.uk}{dhf3@st-andrews.ac.uk}},
Ren\'e Heller$^{3}$ and Michael Hippke$^{4}$
\vspace{0.2cm} \\
$^1$SUPA, School of Physics and Astronomy, University of St Andrews \\
$^2$St Andrews Centre for Exoplanet Science, University of St Andrews \\
$^3$Max Planck Institute for Solar System Research, Justus-von-Liebig-Weg 3, 37077 G\"ottingen, Germany \\
$^4$Sonneberg Observatory, Sternwartestr. 32, 96515 Sonneberg, Germany
}
\date{Accepted XXX. Received YYY; in original form ZZZ}

\pubyear{2017}

\begin{document}
\label{firstpage}
\pagerange{\pageref{firstpage}--\pageref{lastpage}}
\maketitle

\begin{abstract}

Upon entering a star system, light sails are subject to both gravitational forces and radiation pressure, and can use both in concert to modify their trajectory. Moreover, stars possess significant magnetic fields, and if the sail is in any way charged, it will feel the Lorentz force also. 

We investigate the dynamics of so-called ``photogravimagnetic assists'' of sailcraft around $\alpha$ Centauri A, a potential first destination \emph{en route} to Proxima Centauri (the goal of the Breakthrough Starshot program). We find that a 10m$^2$ sail with a charge-to-mass-ratio of around 10 $\mu$C/g or higher will need to take account of magnetic field effects during orbital maneouvres. The magnetic field can provide an extra source of deceleration and deflection, and allow capture onto closer orbits around a target star. 

However, flipping the sign of the sailcraft's charge can radically change resulting trajectories, resulting in complex loop-de-loops around magnetic field lines and essentially random ejection from the star system. Even on well-behaved trajectories, the field can generate off-axis deflections at $\alpha$ Centauri that, while minor, can result in very poor targetting of the final destination (Proxima) post-assist. 

Fortunately for Breakthrough Starshot, nanosails are less prone to charging \emph{en route} than their heavier counterparts, but can still accrue relatively high charge at both the origin and destination, when travelling at low speeds.   Photogravimagnetic assists are highly non-trivial, and require careful course correction to mitigate against unwanted changes in trajectory.

\end{abstract}




\section{Introduction}\label{sec:intro}

\noindent It has been clear since the beginning of the Space Age that effective interstellar travel is impossible using chemical rocket technology.  Our most distant probes achieve maximum velocities of order tens of kilometres per second - the transit time to the nearest star, Proxima Centauri, is measured in hundreds of thousands of years \citep{Wertheimer2006,Kervella2017}.

The recent detection of a close-to Earth mass planet in the radiative liquid water zone around Proxima \citep{Anglada-Escude2016} fires the imagination, and raises the possibility that the star system next door could play host to a viable biosphere \citep{Barnes2016,Meadows2016, Ribas2016,Turbet2016,Dong2017}.  Assays of Proxima b's habitability from Earth will be challenging, as the planet does not appear to transit the star \citep{Kipping2017}.  Atmospheric characterisation via direct imaging and phase curve variations (either in reflected starlight or thermal emission) remain possible in principle with the James Webb Space Telescope \citep{Kreidberg2016}.  

Despite our best efforts, it may be the case that the possibility of life on Proxima b can only be confirmed by \emph{in situ} measurements \citep{Crawford2017}.  If this is so, an interstellar mission whose travel time is measured in years, not millenia or megayears, is essential.

Nuclear propulsion designs, such as the external pulse design \citep{Everett1955} made famous by Project Orion \citep{Mallove1989} or the internal fusion generator design of Daedalus (and Icarus) (e.g. \citealt{Bond1978}) generate significantly higher exhaust velocities than current rocket technology, and can thus achieve much greater cruise velocities, ranging from 0.03-0.1$c$, where $c$ is the speed of light \emph{in vacuo} \citep{Matloff2006}.  Both designs still require a significant proportion of fuel to be placed aboard the craft, even with ramjet technologies \citep{Bussard1960,Bond1974}, not to mention the legal barriers to detonating small fission bombs in space, and the confinement problem in thermonuclear fusion generators.  If antimatter could be synthesised in large quantities, then matter-antimatter annihilation is an extremely powerful energy source from which to develop thrust, but the large-scale synthesis of such fuel currently remains in the realms of science fiction.

Light sails have long been proposed as an energy efficient alternative to matter-based propulsion systems.  Employing the radiation pressure generated by photons, light sails can undergo weak but continuous acceleration to high velocities (for low-mass spacecraft).  The source of photons can either be the host star (solar sails, \citealt{Tsander1961}) or powerful, highly collimated lasers (laser sails, \citealt{Marx1966}).

\citet{Forward1984} described what we now call a ``heavy sail'' concept for the laser sail.  A one metric ton lightsail is illuminated by a laser focused by a 1000 km diameter Fresnel zone lens.  The laserlight provides a constant acceleration of 0.36 $\rm{m \, s^{-2}}$ over a three year period, attaining a cruise velocity of $0.11c$.  

Technological advances, both in the miniaturisation of electronics and the materials science regarding the sail fabric permit the construction of extremely light sailcraft with high cruising velocities.  The Breakthrough Starshot program is considering such a ``nanosail'' approach, designing a ground based laser from a phased array of transmitters, and a factory producing a number of 1g ``StarChip'' sailcraft, which can each be accelerated to velocities around $0.1 - 0.2c$, reaching Proxima in 20-40 years.  When this speed is achieved, waiting for further technological innovation will not get us to Proxima at an earlier date \citep{Heller2017b}.

While such acceleration is extremely beneficial at the beginning of an interstellar mission, it has a corresponding cost at the destination.  Sailcraft travelling at $0.2c$ will traverse an astronomical unit (AU) in around 40 minutes, and the orbit of the Moon around the Earth (0.00257 AU) in around six seconds, reducing the mission's ability to carry out scientific investigations of the Proxima system.

Ideally, an interstellar mission profile would include a deceleration phase, so that the craft can remain in the Proxima system for longer and conduct a fuller investigation, and transmit data to Earth.  For propulsion methods, this requires substantially more fuel, and reduces the maximum cruise velocity as a result.  Light sails carry minimal fuel reserves, so that the sail can be re-oriented, and must rely on local radiation fields to produce significant $\Delta v$\footnote{It is worth noting \citet{Forward1984}'s ingenious scheme for deceleration of laser sails using the Earthbound laser. This is achieved by separating the craft into two sails, to reflect the photons against the direction of travel and decelerate.  This two sail deceleration requires very careful station keeping between both parts, and the laser light cannot be adjusted without several years of time lag.}.

\citet{Heller2017} demonstrated that lightsails could use the destination star system as a source of photons to decelerate.  Combining this photon deceleration with the gravitational force exerted by the star, they showed that a sail travelling at $0.046c$ could achieve a parking orbit around Proxima by using what they dub ``photogravitational assists'' from $\alpha$ Centauri A and B.  Further tuning of the trajectory can reduce the travel time between $\alpha$ Cen A and Proxima from 95 yr to 75 yr \citep{Heller2017a}.

Radiation pressure and gravity are not the only forces acting on a sail as it approaches a star system.  If a sail has accumulated a net charge, then the stellar magnetic field will also exert a force on the sailcraft.  If this charge results in a current, then the craft's own magnetic field may be used as a decelerant \citep{Zubrin1991, Matloff2009, Freeland2015}.  In this paper, we consider \emph{photogravimagnetic} (PGM) assists, where the dynamical evolution of a sail depends on the combination and interplay of radiation, gravity and magnetic forces.  Section \ref{sec:methods} describes the form of these three forces, and the simulations used to explore sail trajectories; section \ref{sec:results} shows some benefits (and costs) from allowing the craft to be charged and using the magnetic field as an extra force; section \ref{sec:discussion}.


\section{Methods} \label{sec:methods}

\noindent We expand on the calculations of \citet{Heller2017}.  The magnetic force is inherently 3D, and as such we will be required to generalise their 2D equations, which we describe below.  Throughout, we define the sail as a craft of mass $M$, and sail area $A$, with the sail's orientation defined by the unit normal vector $\unitnormal$.  The sail is located at position $\mathbf{r}$ relative to the star, with velocity $\mathbf{v}$.

\subsection{Radiation Pressure Force}

\noindent The radiation pressure measured at distance $r$ from a star, luminosity $L_*$, radius $R_*$ is \citep{McInnes1990,Heller2017}:

\begin{equation}
P(r) = \frac{L_*}{3 \pi c R^2_*} \left(1 - \left(1 - \left(\frac{R_*}{r}\right)^2\right)^{3/2} \right),
\end{equation} 

\noindent where we have assumed a uniformly bright finite stellar disc.  The radiation pressure force is then

\begin{equation}
\mathbf{F}_{\rm rad}(\mathbf{r},\unitnormal) = P(r) A \left(\mathbf{\hat{r}} \cdot \unitnormal\right) \unitnormal,
\end{equation}

\noindent and achieves its maximum magnitude when $\unitnormal$ is parallel to $\mathbf{r}$. 

\subsection{Gravitational Force}

\noindent We assume tidal forces on the sail are negligible, and that the sail can be treated as a single pointmass, hence we can use Newton's Law of Gravitation:

\begin{equation}
\mathbf{F}_{\rm grav}(\mathbf{r}) = \frac{-G M_* M}{r^2} \mathbf{\hat{r}}.
\end{equation}

\noindent General relativistic corrections are unlikely to be important, as our sail trajectories have a sufficiently large closest approach distance (for a full general relativistic treatment see \citealt{Kezerashvili2010}).

\subsection{Magnetic Force}

\noindent The force on a sheet with uniform surface charge density $\sigma_q$ moving at velocity $\mathbf{v}$ in a magnetic field $\mathbf{B}$ is given by the Lorentz Force (we assume the electric field $\mathbf{E}=0$ throughout):

\begin{equation}
\mathbf{F} = \sigma_q \int \mathbf{v} \times \mathbf{B} \,d\mathbf{A},
\end{equation}

\noindent where $d\mathbf{A}$ is the surface area element, with direction vector given by $\unitnormal$.  We assume no velocity differential along the sail, and that the magnetic field is uniform across the sail area, and hence we recover the Lorentz force for a point particle with charge $q$:

\begin{equation}
\mathbf{F}_{\rm mag} = \left(\mathbf{v} \times \mathbf{B}\right) \int \sigma_q\, d\mathbf{A} = q \mathbf{v} \times \mathbf{B}.
\end{equation}

\noindent It should be emphasised that $\mathbf{F}_{\rm mag}$ is always a deflective force.  Gravitational and radiative forces can align/anti-align with the velocity, while the Lorentz force is constrained to operate perpendicular to it.  The sign of the charge is also clearly important, as is the orientation of the magnetic field to the approach vector.  These factors can result in both beneficial and hazardous outcomes to a flyby, as we will demonstrate.

\subsection{Optimising Sail Orientation}

\noindent If we wish to maximise the deceleration of our craft due to photon pressure, then we must minimise the component of $\mathbf{F}_{\rm rad}$ along the velocity vector, i.e.:

\begin{equation}
\underset{\unitnormal \in\mathbb{R}^3 }{\text{minimize}} \,\,  \mathbf{F}_{\rm rad}(\mathbf{r},\unitnormal)\cdot \mathbf{v}.
\end{equation}

\noindent We can discard the $P(r)A$ component from $\mathbf{F}_{\rm rad}$ and hence the minimisation problem becomes

\begin{equation}
\underset{\unitnormal \in\mathbb{R}^3 }{\text{minimize}} \,\,  \left(\unitnormal \cdot\mathbf{r}\right)\left(\unitnormal\cdot\mathbf{v}\right).
\end{equation}

\noindent At large distances from the star, the sail will approach the system along a trajectory essentially parallel to its separation vector - hence, $\mathbf{\hat{r}} \cdot\mathbf{\hat{v}}\approx 1$ and we can simply demand that  $\left(\unitnormal\cdot\mathbf{v}\right)$ be minimised, i.e. $\unitnormal \approx -\mathbf{\hat{v}}$.  As we approach the star and velocity decreases, the $\left(\unitnormal\cdot\mathbf{r}\right)$ requirement dominates and $\unitnormal \approx -\mathbf{\hat{r}}$.  

This condition results in a set of linear homogeneous equations that can in principle be solved analytically by Gaussian elimination.  The analytic solutions for this system (in 2D and 3D) is given in Appendix \ref{app:derivation}.  In practice, we use the the sequential least squares algorithm provided by the Python module \texttt{scipy.optimise.minimise} to determine the optimal $\unitnormal$.  

\section{Results} \label{sec:results}

\subsection{When will magnetic fields be important?}

\noindent It is instructive to consider the regimes when the 3 forces will dominate, given fixed stellar properties ($M_*$,$R_*$, $L_*$, $\mathbf{B}_*$).  Throughout this paper, we will assume a pure dipole magnetic field:

\begin{equation}
\mathbf{B}_* = \frac{B_0}{r^3} \left(\frac{\left(\mathbf{\hat{m}}\cdot\mathbf{\hat{r}}\right)\mathbf{\hat{r}}} {r^2} - \mathbf{\hat{m}}\right)
\end{equation}

\noindent Where $\mathbf{\hat{m}}$ is the dipole moment (i.e. the orientation) of the field.  Unless stated otherwise, we tune $B_0$ such that the magnetic field strength at 1 AU resembles that of the solar field measured by Voyager $\sim 5$ nT \citep{Burlaga2002}.  The typical force (in Newtons) will scale as 

\begin{equation}
\left| \mathbf{F}_{\rm mag}\right| = 4 \times 10^{-4} \left(\frac{q}{1\, \rm{\mu C}} \right) \left(\frac{v}{1000\, \kms}\right) \left(\frac{B_0}{5 \, \rm{nT}}\right)\left( \frac{r}{5\, \rm{\rsol}}\right)^{-3} N
\end{equation}

\begin{figure*}
\begin{center}
$\begin{array}{c}
\includegraphics[scale=0.5]{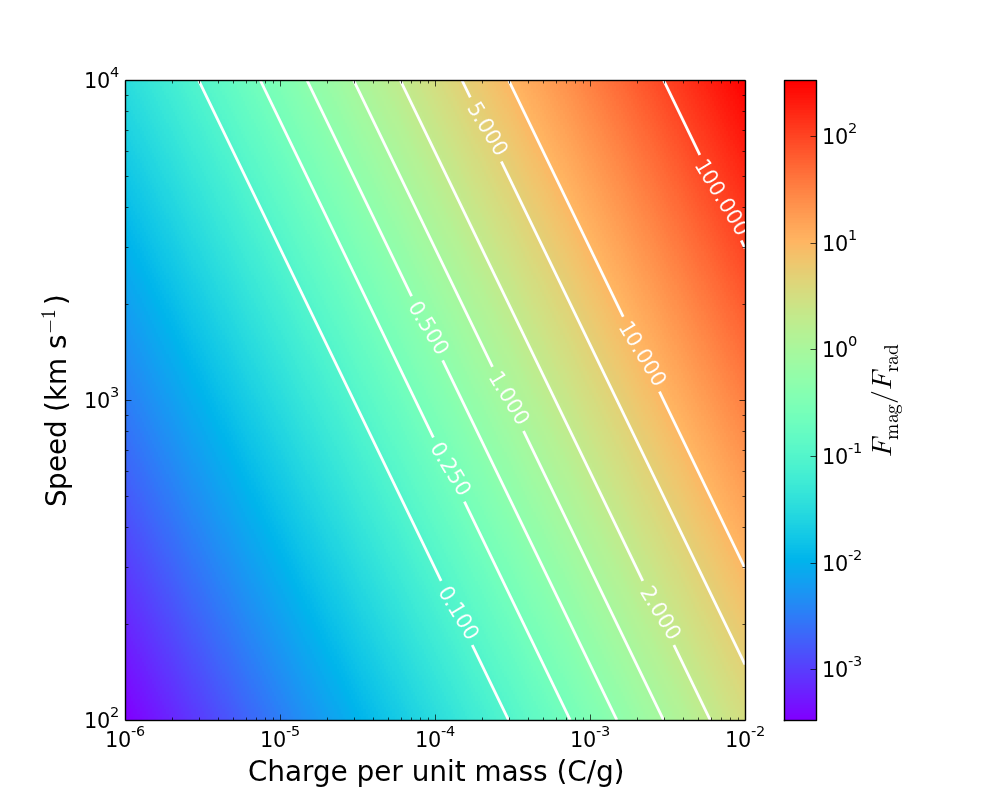} \\
\includegraphics[scale=0.5]{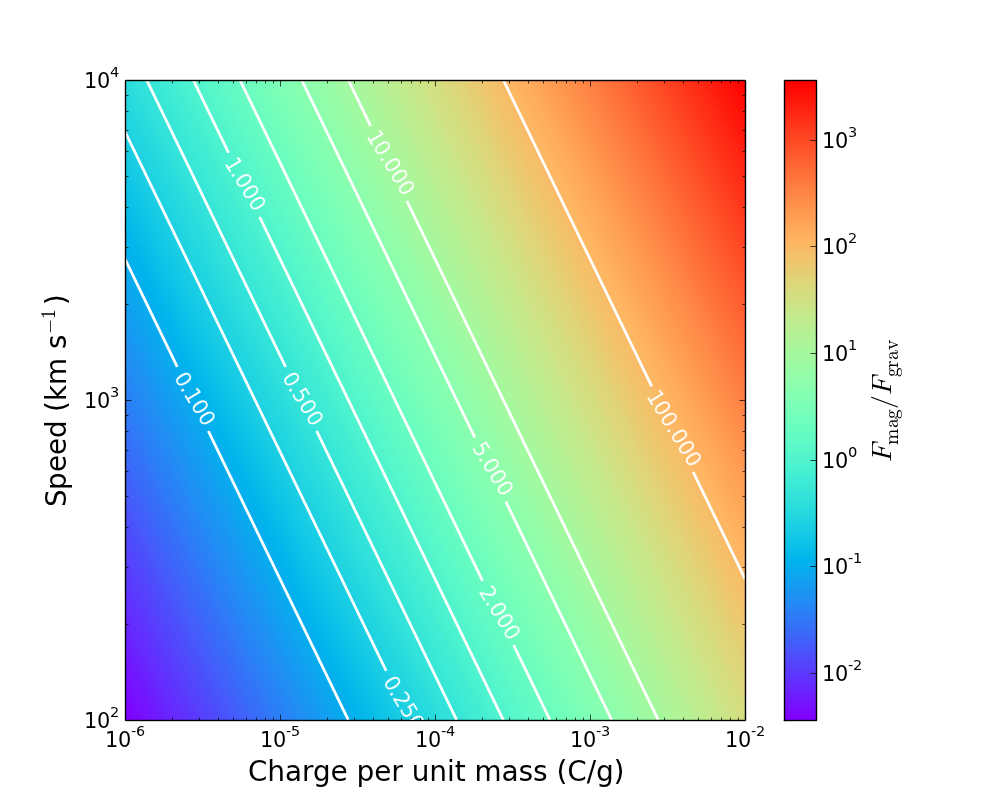} \\
\end{array}$
\end{center}
\caption{The relative strengths of the gravitational, radiation and magnetic forces on a sail, area 10 m$^2$, at a distance of 5 solar radii from the Sun, as a function of its charge per unit mass and speed.  In both cases it is clear that craft with more than 10 $\mu$C/g of charge per unit mass will experience strong magnetic forces.}
\label{fig:compare_forces}
\end{figure*}

\noindent  In Figure \ref{fig:compare_forces} we compute $F_{\rm mag}/F_{\rm rad}$ and $F_{\rm mag}/F_{\rm grav}$ at 5 $\rsol$ from the Sun.  The magnetic moment of the Sun is fixed at $\mathbf{\hat{m}}=(0,0,1)$, and the velocity vector $\mathbf{v}$ is parallel with the negative y-axis.

As the charge on the sail craft exceeds 10 $\mu$ C per gram, the magnetic force becomes a sizeable fraction of the radiative and gravitational forces, and dominates at charges above 100 $\mu$C per gram, regardless of the sail velocity.   

\subsection{Beneficial effects of photogravimagnetic assists}

\subsubsection{Extra Deceleration\label{sec:extradecel}}

\begin{figure}
\begin{center}
\includegraphics[scale=0.5]{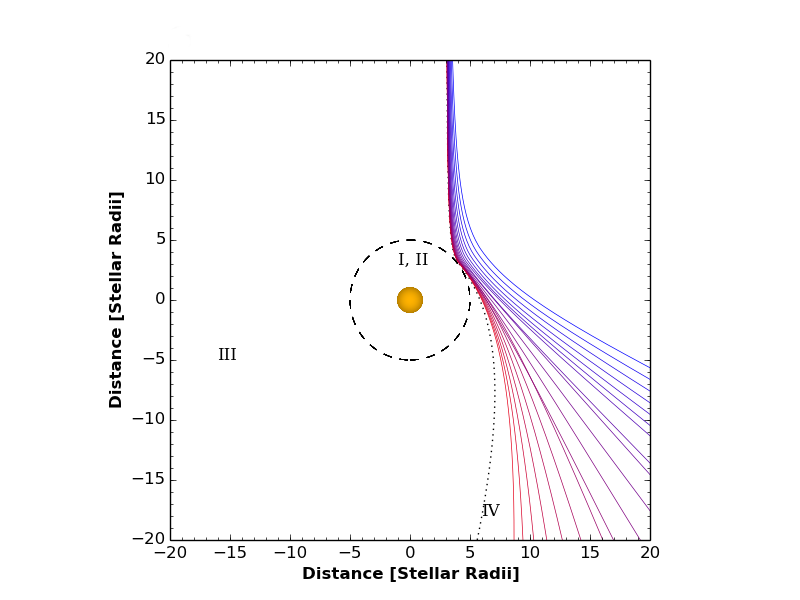}
\end{center}
\caption{The effect of increasing \emph{positive} charge on the photogravimagnetic assist (from 10$\mu C$ (red) to 1 $mC$ (blue)) for all other sail and star parameters held fixed.  The dotted line indicates a zero charge trajectory.  The numbers I-IV indicate the types of possible trajectory described in \citet{Heller2017}: catastrophic loss of the sail (I), full stop (II), bound elliptical orbits (III), and flyby (IV).}
\label{fig:multicharge}
\end{figure}

\noindent Judiciously adding an extra force can result in charged sailcraft undergoing increased deceleration compared to the uncharged case.  As an example, we repeat an example trajectory from \citet{Heller2017} (their Figure 2a).  The sail approaches a star with the mass, luminosity and radius of $\alpha$ Cen A, with magnetic field defined by a dipole with unit moment $\mathbf{\hat{m}}=(0,0,1)$ and a field strength of 5 nT at 1 AU.  The sail's initial velocity vector is

\begin{equation}
\mathbf{v} = (0, -1200 \kms,0),
\end{equation}

\noindent and its initial position vector

\begin{equation}
\mathbf{r} = (3R_*, 10 \rm{AU}, 0).
\end{equation}

\noindent We run a series of flights with increasing positive charge from 10 $\mu$C to 1 $mC$.  The Lorentz force in this case

\begin{equation}
\mathbf{F}_{\rm mag} \propto \frac{\mathbf{v}\times \mathbf{\hat{r}}}{r^5}  - \frac{\mathbf{v}\times \mathbf{\hat{m}}}{r^3} 
\end{equation}

\noindent In this scenario, the first term on the right hand side is directed along the positive z-axis, and the second term along the positive x-axis (with the second term dominating).  Figure \ref{fig:multicharge} shows that an increasing charge boosts this deflection in the x-axis significantly, resulting in a deflection of $\delta = -64^\circ$ for a charge of 1 mC, as opposed to $\delta=15^\circ$ for the uncharged case (dotted line in Figure \ref{fig:multicharge}). 

\subsubsection{Closer Close Approaches}

\begin{figure}
\begin{center}
\includegraphics[scale=0.5]{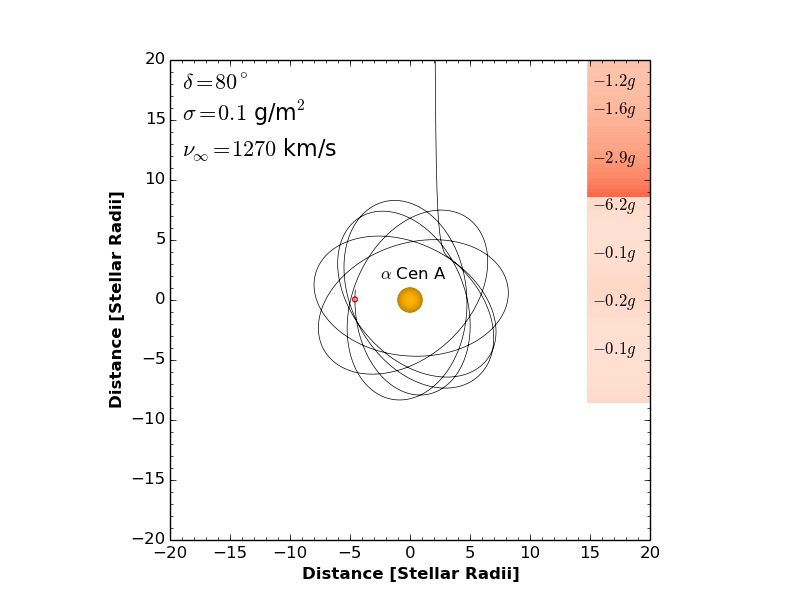}
\end{center}
\caption{Orbit injection at closer approaches thanks to the Lorentz force.  A trajectory that would result in collision with the star for chargeless sails becomes an elliptical orbit for $q=0.1$ mC. \label{fig:spirograph}}
\end{figure}

\noindent Figure \ref{fig:spirograph} shows a similar approach to the previous section, but we now decrease the initial $x$ from $3R_*$ to $2R_*$.  In the absence of magnetic forces the sail would experience a catastrophic deceleration, resulting in its impacting the star.  Adding a positive charge of 0.1 mC allows the craft to be injected into a highly elliptic, precessing orbit (Figure \ref{fig:spirograph}).

This precession could be corrected with judicious angling of the sail, which we have not attempted here.  Ejection from the system on a desired trajectory could also be triggered with appropriate tacking of the sail (in principle).

\subsection{Detrimental Effects of photogravimagnetic assists}

\subsubsection{Growing $z$ offsets in trajectories}

\begin{figure}
\begin{center}
\includegraphics[scale=0.5]{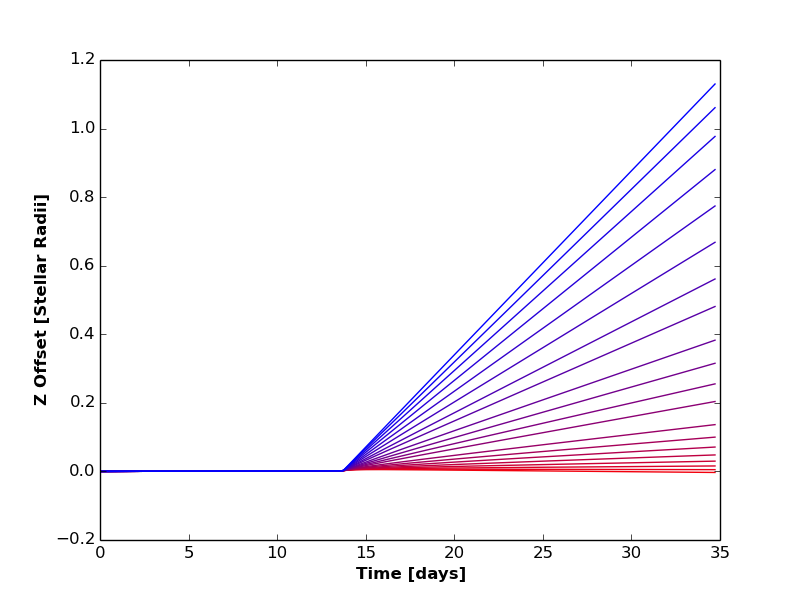}
\end{center}
\caption{Growing $z$ offsets in trajectories as (positive) charge is increased from 10$\mu C$ (red) to 1 $mC$ (blue). The sailcrafts' distance from the star approximately 3 days post-assist ranges between 1 and 5 AU depending on charge.  \label{fig:z_offset}}
\end{figure}

\noindent We are unlikely to achieve precisely 2D trajectories on approach to $\alpha$ Centauri (i.e. $z=0$, $v_z=0$).  Even small offsets in either parameter is present as we enter the star system, we find that it can be significantly amplified on approach.  In Figure \ref{fig:z_offset} we plot the evolution of the $z$ coordinate for the runs displayed in section \ref{sec:extradecel}, with $z=1$ cm, and $v_z=1 \cms$.  As positive charge is increased the $z$ offset can reach a few stellar radii at distances of 1-5 AU from the target (i.e. an angular offset of up to 3 arcminutes).  This $z$ offset will continue to grow as the sail moves towards its intended target of Proxima, most likely resulting in missing its target by tens to hundreds of AU depending on the charge.

\subsubsection{Photogravimagnetic loop-de-loops}

\begin{figure}
\begin{center}
\includegraphics[scale=0.5]{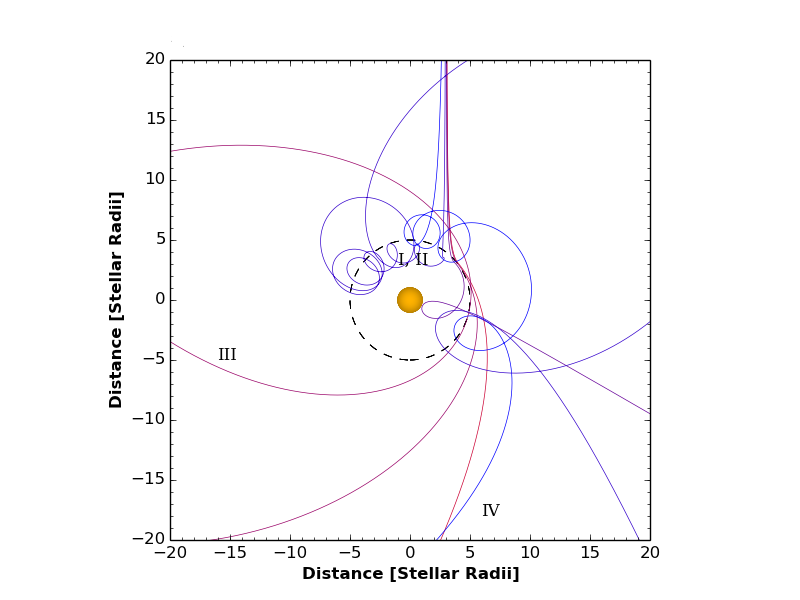}
\end{center}
\caption{The effect of increasing \emph{negative} charge on the photogravimagnetic assist (from 10$\mu C$ (red) to 1 $mC$ (blue)) for all other sail and star parameters held fixed. \label{fig:loopdeloop}}
\end{figure}

\noindent The awkward nature of PGM assists becomes clear when we repeat the set of flights in section \ref{sec:extradecel}, but instead allow the charge to be negative as opposed to positive.  The $x$-component of $F_{\rm mag}$ becomes negative, allowing for some complicated loop-de-loop trajectories as the sail begins to orbit the field lines (Figure \ref{fig:loopdeloop}).   The sail experiences large accelerations, which grow as the sail's speed increases.  The final trajectory of the sail is extremely difficult to predict, and indeed may be ejected at a higher velocity than which it entered the system.

Depending on the mission profile, this extra acceleration might be a benefit, especially if the target is a more distant star, or Galactic exploration is the goal (cf \citealt{Nicholson2013}).  However, achieving an ejection from the system on a desired trajectory is highly non-trivial.  In our case, we wish to decelerate the craft, so it is immediately clear that for this injection geometry a negative charge is best avoided.  If the craft's initial trajectory was flipped in the $y$-axis (i.e. $x\rightarrow -x$), or the magnetic field's dipole moment was reversed, we would reproduce the better behaved trajectories seen in section \ref{sec:extradecel}.

\subsubsection{Large variances in possible trajectories on approach}

\noindent Given that we are limited in our ability to specify the approach vector of the sailcraft at parsec distances from Earth, how does this affect our ability to place the sailcraft on a desired trajectory after a PGM assist? The properties of our target star are likely to fluctuate with time, particularly the luminosity, and magnetic field strength/orientation.

To answer this question, we execute a series of flights where we randomly sample sail and star properties from uniform distributions.  We allow the magnetic field strength to vary between $B_0 = [2.5,7.5]$ nT to reflect the evolution of field strength over the 12-15 yr magnetic cycle of $\alpha$ Cen A \citep{Robrade2012}.  We also allow the field orientation to vary via rotations in the $x$ and $z$ axis by angles between [-5,5]$^\circ$.  The Sun's luminosity only varies by around 0.1\% over these cycles \citep{Krivova2007}, which we will also assume for $\alpha$ Cen A. Hence we randomly sample luminosities in the range [0.999,1.001]$L_*$ .

We allow the sail's charge to vary between 0 and 1 mC, and the initial position vector is

\begin{equation}
\mathbf{r} = \left(x R_*, 10 \rm{AU},0.0\right)
\end{equation}

\begin{figure*}
\begin{center}
$\begin{array}{cc}
\includegraphics[scale=0.35]{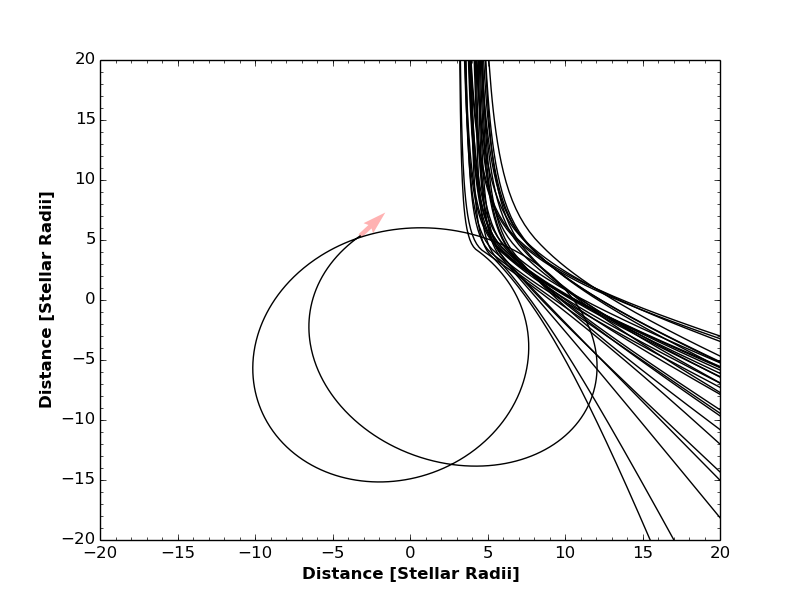} &
\includegraphics[scale=0.35]{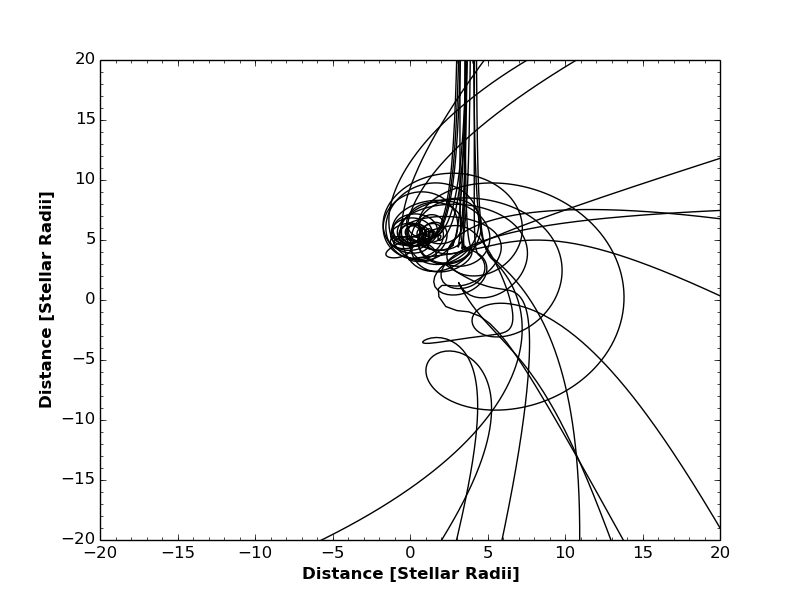} \\
\end{array}$
\end{center}
\caption{Variance in sail trajectories due to uncertainties in stellar and sail parameters.  Left: possible trajectories for positively charged sailcraft.  Right: possible trajectories for negatively charged sailcraft.}
\label{fig:randomsample}
\end{figure*}

\noindent With $x=[2.5-3.5]$.  Figure \ref{fig:randomsample} shows the results of 15 flights, with either positive charge (left panel) or negative charge (right panel).  In both cases, we can see that the final outcome of the close approach to $\alpha$ Cen A can vary widely, between capture into an elliptical orbit, a deceleration and deflection (the intended aim), and highly erratic trajectories with almost random ejection vectors (in the negatively charged case).

\section{Discussion}\label{sec:discussion}

\subsection{What is the expected charge on a sail?}

\noindent There are various natural surface charging mechanisms that our sail will encounter on its route to Proxima (see for example \citealt{Garrett2012}).  We identify two classes of environment in which charging should be considered - the first is in the vicinity of the Sun (or Proxima), and the second is in the interstellar medium between both stars.

\subsubsection{Charging in the Solar system/at Proxima}

In geosynchronous Earth orbit (GEO), surface charging may occur where GEO exits the Earth's plasmasphere \citep{Mikaelian2009}.  This depends on the Sun's activity and the geomagnetic response, but in the worst case scenario a spacecraft can be bombarded with high energy electrons with no protection from the plasmasphere, and undergo strong charging.  If the craft is illuminated by the Sun these effects can be ameliorated by photoelectron emission.

On the other hand, photoelectron emission can also generate strong positive charges.  As most of our flight plan relies on the collection of large quantities of photons, we must be certain that our sail is not being overly charged by photoelectron production.  This essentially relies on low absorption levels, which is already a stringent technical requirement of any viable Starshot mission.  Nanosail designs hope to achieve absorption levels of 10 parts per million or less \citep{Lubin2016}. 

Most studies of spacecraft charging quote the electric potential generated.  An object immersed in a plasma will achieve a floating potential given by

\begin{equation}
V_{f} = -\frac{k_{B}T_{e}}{2e}\ln\left(\frac{m_{i}}{2 \pi m_{e}}\right),
\end{equation}

\noindent where $T_e$ is the electron temperature, $k_{B}$ is the Boltzmann constant, $m_i$ and $m_e$ are the ion and electron mass respectively, and $e$ is the electron charge.  The potential can be as large as $V_f=20 kV$ for spacecraft in orbit around the Earth depending on which process is doing the charging \citep{Mikaelian2009}.

A simple estimate for a spherical object of radius $L$ suggests that the typical charge in this case would be

\begin{equation}
q = 4\pi \epsilon_0 V L = 2.2 \times 10^{-6} \left(\frac{V_f}{20 \,\rm{kV}}\right) \left(\frac{L}{ 1\, \rm{m}}\right)\, \rm{C}.
\end{equation}

\noindent We can therefore expect charges of order a few $\mu$ C for a sail of length/width $L=1m$ (at least near to the Earth).  $V_f$ increases linearly with electron temperature.  In the solar wind, this temperature is around $10^4-10^5$ K, but increases by an order of magnitude beyond the solar termination shock, accompanied by strong fluctuations in $T_e$ across the heliosheath \citep{Richardson2008,Chashei2013}.  These estimates suggest that our sail may already achieve sufficient charge for Lorentz forces to be important before exiting the Solar system, although it is important to note that the above equation is only valid for when the craft is moving at velocities much lower than the electron thermal velocity given by $T_e$.

We can therefore expect that during the pre-launch phase, a 10$m^2$ sail may be charged up to of order 10$\mu$C if it is initially released outside the Earth's plasmasphere.  Once the sail is accelerated to relativistic speeds, the charging timescale becomes long compared to the interaction time of the sail with the plasma, so subsequent charging as the sail leaves the Solar system is reduced.  Equally, as the craft decelerates on entry to the Proxima system, the interaction time increases until it is (likely) comparable to the charging time.  It is therefore at the very beginning and end of the mission that charging becomes important.

\subsubsection{Charging in the ISM}

\noindent As the sail traverses the interstellar medium, it will encounter a range of different charged particles, from electrons to highly porous dust grains.  The dominant charging process for the sail is electron bombardment \citep{Hoang2017a}, as the energy density of ultraviolet photons in the ISM is too low for photoelectron emission.

One could imagine the sail collecting a layer of charged dust grains in transit to Proxima.  This would not only increase the Lorentz force on the craft, but also its absorption coefficient, with potentially fatal results.  \citet{Lubin2016} estimate the total dust column along the sail trajectory to be approximately $4 \times 10^7 \, m^{-2}$ (for grain sizes greater than a micron).  An edge on sailcraft can expect of order a thousand grain collisions (depending on its thickness), suggesting a relatively weak charge.  At relativistic speeds, grain collisions are unlikely to result in sticking, but instead piercing the sail.  This results in little to no charging, and fortunately little effect on the sail velocity, but can introduce inhomogeneities in the sail, reducing the sail's ability to produce stable flight \citep{Hoang2017}, or in the case of dust grains larger than 15 $\mu$m, complete destruction of the spacecraft \citep{Hoang2017b}.

At relativistic speeds, the maximum potential achieved by the spacecraft depends on its speed, not the plasma's properties.  Consequently, the spacecraft will not achieve the floating potential of the plasma, and assume a much lower value, typically less than a $\mu$V per gram \citep{Hoang2017}.

\subsubsection{So what is the charge?}

In the Solar system, simple energy arguments would indicate that due to their lower mass, electron charging is more likely than ionic charging, suggesting a typically negative charge.  However, \emph{en route} to Proxima, collisional emission of electrons is likely to generate a positive charge.

Throughout, we have discussed charge in absolute terms, and assumed that the spacecraft charge density was uniform.  Differential or relative charging is likely to occur, especially when the craft is composed of materials of varying dielectric properties.  If the surface charge density of our sail is non-uniform, the Lorentz force will introduce a stress across the sail surface.  Weak stresses can result in tumbling motions as regions of the sail are accelerated at different rates, which can result in rotation periods as short as $\sim$ 30 minutes \citep{Hoang2017a}.  If the stress exceeds the strength of the sail material, this can result in irreparable damage.  

The \emph{initial} total charge of our sail (generated before the craft is accelerated) seems to depend most strongly on the space environment in the immediate region of the Sun, and similarly at its destination as the craft decelerates to non-relativistic velocities.  It seems to be the case that the sail may achieve a sufficiently large charge for photogravimagnetic assists to be important at the beginning of the mission, but that the net charge will be reduced as it traverses the ISM.

However, we still lack detailed knowledge of the charge content of the intervening ISM along our proposed flight plan.  Also, it is clear that the ambient magnetic field plays a role in trapping surface charge via the Lorentz force, or allowing it to stream away from the craft along field lines.  Most concerningly, all this implies that the surface charge is likely to be a function of not only distance from Proxima, but also the sail normal on approach!

\subsection{What is the preferred charge on a sail?}

\noindent We have seen that the sign of the charge affects its resulting trajectory profoundly.  This change is entirely grounded in the sign of the magnetic force.  If the magnetic force acts away from the star, then a stable trajectory is obtained (like those seen in the left panel of Figure \ref{fig:randomsample}).  If the magnetic force is attractive, then we recover erratic trajectories such as those seen in the right panel of Figure \ref{fig:randomsample}.

The preferred charge depends on how the sail approaches the star.  We have seen negative charges providing hazardous trajectories because the sail approaches the star in the positive $x$-axis, positive $y$-axis quadrant.  If the flight is carried out in the negative $x$, positive $y$ quadrant, then a positive charge produces loop-de-loop trajectories.

If we want to avoid erratic flights, we should therefore be aiming to either a) approach the star in such a way that the magnetic force is repulsive, or b) charge the sail in such a way that the magnetic force is repulsive.  If we know what the charge is expected to be (and its sign) from natural processes, we can adjust our ingress trajectory to ensure that the egress trajectory is well-behaved, or we can deliberately charge the sail to achieve the same effect.

What methods are there for charging sails? Given that the most likely source of charging \emph{en route} is emission of electrons as a result of particle bombardment \citep{Hoang2017a}, we should expect that our craft will typically be positively charged.  If the craft is negatively charged, orienting the sail to stimulate photoelectron emission (if we are near a strong UV source) can reduce an accrued negative charge.  Reducing a positive charge would mostly likely require collection of electrons in transit to Proxima, and is unlikely to be achievable.  Electrons will pass through the sail as their penetration length is long compared to the sail thickness.  The sail would require some kind of charge capture system that could decelerate the electrons sufficiently for capture (such as a magnetic field generator of its own).

\subsection{Limitations of the Analysis}

\noindent We have focused entirely on one optimisation scheme for sail orientation (which maximises the component of force antiparallel to velocity).  It is quite possible to implement other schemes, such as using the sail to attempt to cancel the magnetic force contribution, i.e.

\begin{equation}
\underset{\unitnormal \in\mathbb{R}^3 }{\text{minimize}} \,\,  \mathbf{F}_{\rm rad}(\mathbf{r},\unitnormal).\mathbf{F}_{\rm mag}.
\end{equation}

\noindent which would mitigate against more erratic trajectories, but does not guarantee maximal deceleration during the encounter.  

We have also focused largely on pure dipole magnetic fields, where all field lines are entirely closed.  This is a heavy oversimplification of realistic stellar magnetic fields, which typically contain higher order spherical harmonics than the pure dipole, and also contain open field lines at large distance from the star.  A probe traversing one of these ``real'' magnetic fields might feel very little Lorentz force until it reaches the region where field lines begin to close.  

The configuration of stellar magnetic fields also evolves with time.  Magnetic reconnection and coronal mass ejections are common around G stars like $\alpha$ Centauri A, and flaring is a key characteristic of low mass stars such as Proxima.  These eruptive events will add extra time dependence to the magnetic field, and hence the Lorentz force experienced by the sail.  Of course, a sail that directly encounters a coronal mass ejection during close approach will most likely be damaged beyond repair.

We have also ignored magnetic fields in the ISM \emph{en route}.  If the level of ionisation is sufficiently high, magnetic fields tend to ``freeze in'' to the turbulent density structures in the diffuse gas.  At large scales, this turbulent structure is a complex mix of filaments, sheets and knots, with the field tending to align with these structures except in regions of poor ionisation \citep{McClure-Griffiths2006,Begum2010,Clark2014}.  We should therefore expect that our sailcraft will be experiencing Lorentz forces along its entire flight path.  The direction of this force will be extremely sensitive to the density structure of the ISM it passes through.  Further work is needed to determine the level of deflection a craft might experience as it traverses a turbulent magnetic field environment, and what level of charge mitigation is required to avoid substantial changes in trajectory \citep[cf][]{Gros2017,Hoang2017a}.

Finally, we have ignored the more ``standard'' uses of electromagnetic fields in sail navigation.  Indeed, the literature is replete with works on ``magsails'' that generate their own fields to deflect either interstellar ions \citep{Zubrin1991,Freeland2015} or charged particles entrained in the stellar wind \citep{Matloff2009}.  This produces a drag force on the craft, providing an effective means of deceleration.  The magnitude of the deceleration depends on the effective field geometry, in particular its cross-section to the wind-flow, as well as the sail's velocity relative to the stellar wind.  Electric sails are composed of tethers deliberately charged to maintain a high positive voltage, deflecting high energy protons \citep{Janhunen2004}.  A two stage magsail / electric sail combination has been shown to be an effective deceleration strategy for high mass spacecraft \citep{Perakis2016}.  Electric sails require high voltage (and possibly significant charging), and it is unclear how this would interact with Lorentz forces.  On the other hand, generating a magnetic field may be useful for mitigating excess charging and permitting charge escape.  Combining photogravimagnetic assists with stellar wind deceleration may provide a highly effective means of generating $\Delta v$, and should be explored further.

\section{Conclusions}\label{sec:conclusion}

\noindent We have simulated the trajectories of light sails under the combined forces of gravity, radiation pressure and magnetic fields.  If wafer sails, such as those proposed by the Breakthrough Starshot program, obtain a greater charge than around 10 $\mu C$ (per gram of mass) during flight, magnetic fields begin to significantly affect their final trajectory. 

The Lorentz force generated by the magnetic field can be either beneficial or hazardous depending on the inbound trajectory.  In the best case, the sail can use photogravimagnetic (PGM) assists to generate even greater $\Delta v$ than achievable with only gravity and radiation.  This allows for craft to perform stellar flybys at even closer approaches, and to attain a higher cruise velocity on leaving Earth and still be able to achieve a successful flyby of Proxima.

On the other hand, the worst case scenario can result in highly erratic loop-de-loop trajectories, which either result in impact onto the star or ejection at high speed on highly unpredictable outbound trajectories.  Even if these trajectories can be avoided, small changes in trajectory at ingress can be multiplied into large offsets at egress, requiring large course corrections if our sail is to reach its intended target.  Deflections can even be generated on departure by the Sun's magnetic field, if the craft is charged during launch.

The contingent nature of these photogravimagnetic trajectories suggests that sail missions should be prepared to send multiple craft to boost their probability of executing a PGM assist that delivers a preferred trajectory - that is, a trajectory towards Proxima at relatively low velocity.

Photogravimagnetic assists are something of a mixed blessing, and their effects on navigation must be considered carefully.  We recommend that planning for Breakthrough Starshot commits resources to a) monitoring sailcraft charging and Lorentz force strengths during Solar system tests, b) mapping sources of sailcraft charging between the Solar system and $\alpha$ Centauri, and c) mapping the magnetic field of $\alpha$ Cen A, B and C, as well as the interstellar environment along the route.

\section*{Acknowledgments}

DF gratefully acknowledges support from the ECOGAL project, grant agreement 291227, funded by the European Research Council under ERC-2011-ADG.  This work was supported in part by the German space agency (Deutsches Zentrum f\"ur Luft- und Raumfahrt) under PLATO Data Center grant 50OO1501.  The authors warmly thank Craig Stark and the anonymous reviewer for stimulating discussions on spacecraft charging.




\bibliographystyle{mnras} 
\bibliography{photogravmag}

\begin{thebibliography}{}
\makeatletter
\relax
\def\mn@urlcharsother{\let\do\@makeother \do\$\do\&\do\#\do\^\do\_\do\%\do\~}
\def\mn@doi{\begingroup\mn@urlcharsother \@ifnextchar [ {\mn@doi@}
  {\mn@doi@[]}}
\def\mn@doi@[#1]#2{\def\@tempa{#1}\ifx\@tempa\@empty \href
  {http://dx.doi.org/#2} {doi:#2}\else \href {http://dx.doi.org/#2} {#1}\fi
  \endgroup}
\def\mn@eprint#1#2{\mn@eprint@#1:#2::\@nil}
\def\mn@eprint@arXiv#1{\href {http://arxiv.org/abs/#1} {{\tt arXiv:#1}}}
\def\mn@eprint@dblp#1{\href {http://dblp.uni-trier.de/rec/bibtex/#1.xml}
  {dblp:#1}}
\def\mn@eprint@#1:#2:#3:#4\@nil{\def\@tempa {#1}\def\@tempb {#2}\def\@tempc
  {#3}\ifx \@tempc \@empty \let \@tempc \@tempb \let \@tempb \@tempa \fi \ifx
  \@tempb \@empty \def\@tempb {arXiv}\fi \@ifundefined
  {mn@eprint@\@tempb}{\@tempb:\@tempc}{\expandafter \expandafter \csname
  mn@eprint@\@tempb\endcsname \expandafter{\@tempc}}}

\bibitem[\protect\citeauthoryear{Anglada-Escud{\'{e}}
  et~al.,}{Anglada-Escud{\'{e}} et~al.}{2016}]{Anglada-Escude2016}
Anglada-Escud{\'{e}} G.,  et~al., 2016, \mn@doi [Nature] {10.1038/nature19106},
  536, 437

\bibitem[\protect\citeauthoryear{Barnes et~al.,}{Barnes
  et~al.}{2016}]{Barnes2016}
Barnes R.,  et~al., 2016, eprint arXiv:1608.06919

\bibitem[\protect\citeauthoryear{Begum et~al.,}{Begum et~al.}{2010}]{Begum2010}
Begum A.,  et~al., 2010, \mn@doi [The Astrophysical Journal]
  {10.1088/0004-637X/722/1/395}, 722, 395

\bibitem[\protect\citeauthoryear{Bond}{Bond}{1974}]{Bond1974}
Bond A.,  1974, Journal of the British Interplanetary Society, 27, 674

\bibitem[\protect\citeauthoryear{Bond}{Bond}{1978}]{Bond1978}
Bond A.,  1978, {Project Daedalus: The Final Report on the BIS Starship Study}.
British Interplanetary Society

\bibitem[\protect\citeauthoryear{Burlaga}{Burlaga}{2002}]{Burlaga2002}
Burlaga L.~F.,  2002, \mn@doi [Journal of Geophysical Research]
  {10.1029/2001JA009217}, 107, 1410

\bibitem[\protect\citeauthoryear{Bussard}{Bussard}{1960}]{Bussard1960}
Bussard R.,  1960, Acta Astronautica, 6, 170

\bibitem[\protect\citeauthoryear{Chashei \& Fahr}{Chashei \&
  Fahr}{2013}]{Chashei2013}
Chashei I.~V.,  Fahr H.~J.,  2013, \mn@doi [Annales Geophysicae]
  {10.5194/angeo-31-1205-2013}, 31, 1205

\bibitem[\protect\citeauthoryear{Clark, Peek  \& Putman}{Clark
  et~al.}{2014}]{Clark2014}
Clark S.~E.,  Peek J. E.~G.,   Putman M.~E.,  2014, \mn@doi [The Astrophysical
  Journal] {10.1088/0004-637X/789/1/82}, 789, 82

\bibitem[\protect\citeauthoryear{Crawford}{Crawford}{2017}]{Crawford2017}
Crawford I.~A.,  2017, in Deeg H.,  Bolmonte J.,  eds, , Handbook of
  Exoplanets.
Springer (\mn@eprint {arXiv} {1707.01174}), \url
  {http://arxiv.org/abs/1707.01174}

\bibitem[\protect\citeauthoryear{Dong, Lingam, Ma  \& Cohen}{Dong
  et~al.}{2017}]{Dong2017}
Dong C.,  Lingam M.,  Ma Y.,   Cohen O.,  2017, \mn@doi [ApJ]
  {10.3847/2041-8213/aa6438}, 837, L26

\bibitem[\protect\citeauthoryear{Everett \& Ulam}{Everett \&
  Ulam}{1955}]{Everett1955}
Everett C.~J.,  Ulam S.~M.,  1955, Technical report, {On a Method of Propulsion
  of Projectiles by Means of External Nuclear Explosions. Part 1.}.
LOS ALAMOS SCIENTIFIC LAB ALBUQUERQUE NM

\bibitem[\protect\citeauthoryear{Forward}{Forward}{1984}]{Forward1984}
Forward R.~L.,  1984, \mn@doi [Journal of Spacecraft and Rockets]
  {10.2514/3.8632}, 21, 187

\bibitem[\protect\citeauthoryear{Freeland}{Freeland}{2015}]{Freeland2015}
Freeland R.,  2015, JBIS, 68, 306

\bibitem[\protect\citeauthoryear{Garrett \& Whittlesey}{Garrett \&
  Whittlesey}{2012}]{Garrett2012}
Garrett H.~B.,  Whittlesey A.~C.,  2012, {Guide to Mitigating Spacecraft
  Charging Effects}.
JPL Space Science and Technology Series, Wiley, \url
  {https://books.google.co.uk/books?id=orFDqGO-iM0C}

\bibitem[\protect\citeauthoryear{Gros}{Gros}{2017}]{Gros2017}
Gros C.,  2017, eprint arXiv:1707.02801

\bibitem[\protect\citeauthoryear{Heller}{Heller}{2017}]{Heller2017b}
Heller R.,  2017, \mn@doi [MNRAS] {10.1093/mnras/stx1493}, 470, 3664

\bibitem[\protect\citeauthoryear{Heller \& Hippke}{Heller \&
  Hippke}{2017}]{Heller2017}
Heller R.,  Hippke M.,  2017, \mn@doi [ApJ] {10.3847/2041-8213/835/2/L32}, 835,
  L32

\bibitem[\protect\citeauthoryear{Heller, Hippke  \& Kervella}{Heller
  et~al.}{2017}]{Heller2017a}
Heller R.,  Hippke M.,   Kervella P.,  2017, \mn@doi [The Astronomical Journal]
  {10.3847/1538-3881/aa813f}, 154, 115

\bibitem[\protect\citeauthoryear{Hoang}{Hoang}{2017}]{Hoang2017}
Hoang T.,  2017, ApJ, p. in press

\bibitem[\protect\citeauthoryear{Hoang \& Loeb}{Hoang \&
  Loeb}{2017}]{Hoang2017a}
Hoang T.,  Loeb A.,  2017, ApJ, p. in press

\bibitem[\protect\citeauthoryear{Hoang, Lazarian, Burkhart  \& Loeb}{Hoang
  et~al.}{2017}]{Hoang2017b}
Hoang T.,  Lazarian A.,  Burkhart B.,   Loeb A.,  2017, \mn@doi [The
  Astrophysical Journal] {10.3847/1538-4357/aa5da6}, 837, 5

\bibitem[\protect\citeauthoryear{Janhunen}{Janhunen}{2004}]{Janhunen2004}
Janhunen P.,  2004, \mn@doi [Journal of Propulsion and Power] {10.2514/1.8580},
  20, 763

\bibitem[\protect\citeauthoryear{Kervella, Th{\'{e}}venin  \& Lovis}{Kervella
  et~al.}{2017}]{Kervella2017}
Kervella P.,  Th{\'{e}}venin F.,   Lovis C.,  2017, \mn@doi [Astronomy {\&}
  Astrophysics] {10.1051/0004-6361/201629930}, 598, L7

\bibitem[\protect\citeauthoryear{Kezerashvili \&
  V{\'{a}}zquez-Poritz}{Kezerashvili \&
  V{\'{a}}zquez-Poritz}{2010}]{Kezerashvili2010}
Kezerashvili R.~Y.,  V{\'{a}}zquez-Poritz J.~F.,  2010, \mn@doi [Advances in
  Space Research] {10.1016/j.asr.2010.03.015}, 46, 346

\bibitem[\protect\citeauthoryear{Kipping et~al.,}{Kipping
  et~al.}{2017}]{Kipping2017}
Kipping D.~M.,  et~al., 2017, \mn@doi [The Astronomical Journal]
  {10.3847/1538-3881/153/3/93}, 153, 93

\bibitem[\protect\citeauthoryear{Kreidberg \& Loeb}{Kreidberg \&
  Loeb}{2016}]{Kreidberg2016}
Kreidberg L.,  Loeb A.,  2016, \mn@doi [ApJ] {10.3847/2041-8205/832/1/L12},
  832, L12

\bibitem[\protect\citeauthoryear{Krivova, Balmaceda  \& Solanki}{Krivova
  et~al.}{2007}]{Krivova2007}
Krivova N.~A.,  Balmaceda L.,   Solanki S.~K.,  2007, \mn@doi [Astronomy {\&}
  Astrophysics] {10.1051/0004-6361:20066725}, 467, 335

\bibitem[\protect\citeauthoryear{Lubin}{Lubin}{2016}]{Lubin2016}
Lubin P.,  2016, JBIS, 69, 40

\bibitem[\protect\citeauthoryear{Mallove \& Matloff}{Mallove \&
  Matloff}{1989}]{Mallove1989}
Mallove E.~F.,  Matloff G.~L.,  1989, {The starflight handbook : a pioneer's
  guide to interstellar travel}.
Wiley

\bibitem[\protect\citeauthoryear{Marx}{Marx}{1966}]{Marx1966}
Marx G.,  1966, \mn@doi [Nature] {10.1038/211022a0}, 211, 22

\bibitem[\protect\citeauthoryear{Matloff}{Matloff}{2006}]{Matloff2006}
Matloff G.,  2006, {Deep Space Probes: To the Outer Solar System and Beyond},
  2nd edn.
Springer Science {\&} Business Media

\bibitem[\protect\citeauthoryear{Matloff}{Matloff}{2009}]{Matloff2009}
Matloff G.,  2009, JBIS, 62, 66

\bibitem[\protect\citeauthoryear{McClure-Griffiths}{McClure-Griffiths}{2006}]{%
McClure-Griffiths2006}
McClure-Griffiths N.,  2006, in New Horizons in Astronomy: Frank N. Bash
  Symposium ASP Conference Series. Astronomical Society of the Pacific (ASP),
  p.~95, \url {http://adsabs.harvard.edu/abs/2006ASPC..352...95M}

\bibitem[\protect\citeauthoryear{McInnes \& Brown}{McInnes \&
  Brown}{1990}]{McInnes1990}
McInnes C.~R.,  Brown J.~C.,  1990, \mn@doi [Celestial Mechanics and Dynamical
  Astronomy] {10.1007/BF00049416}, 49, 249

\bibitem[\protect\citeauthoryear{Meadows et~al.,}{Meadows
  et~al.}{2016}]{Meadows2016}
Meadows V.~S.,  et~al., 2016, eprint arXiv:1608.08620

\bibitem[\protect\citeauthoryear{Mikaelian \& Tsoline}{Mikaelian \&
  Tsoline}{2009}]{Mikaelian2009}
Mikaelian T.,  Tsoline 2009, eprint arXiv:0906.3884

\bibitem[\protect\citeauthoryear{Nicholson \& Forgan}{Nicholson \&
  Forgan}{2013}]{Nicholson2013}
Nicholson A.,  Forgan D.,  2013, \mn@doi [International Journal of
  Astrobiology] {10.1017/S1473550413000244}, 12, 337

\bibitem[\protect\citeauthoryear{Perakis \& Hein}{Perakis \&
  Hein}{2016}]{Perakis2016}
Perakis N.,  Hein A.~M.,  2016, \mn@doi [Acta Astronautica]
  {10.1016/j.actaastro.2016.07.005}, 128, 13

\bibitem[\protect\citeauthoryear{Ribas et~al.,}{Ribas et~al.}{2016}]{Ribas2016}
Ribas I.,  et~al., 2016, \mn@doi [Astronomy {\&} Astrophysics]
  {10.1051/0004-6361/201629576}, 596, A111

\bibitem[\protect\citeauthoryear{Richardson}{Richardson}{2008}]{Richardson2008}
Richardson J.~D.,  2008, \mn@doi [Geophysical Research Letters]
  {10.1029/2008GL036168}, 35, L23104

\bibitem[\protect\citeauthoryear{Robrade, Schmitt  \& Favata}{Robrade
  et~al.}{2012}]{Robrade2012}
Robrade J.,  Schmitt J. H. M.~M.,   Favata F.,  2012, \mn@doi [Astronomy {\&}
  Astrophysics] {10.1051/0004-6361/201219046}, 543, A84

\bibitem[\protect\citeauthoryear{Tsander}{Tsander}{1961}]{Tsander1961}
Tsander F.,  1961, {Problems of flight by jet propulsion: interplanetary
  flights}.
Nawzhio - Technical Publishing House, Moscow

\bibitem[\protect\citeauthoryear{Turbet, Leconte, Selsis, Bolmont, Forget,
  Ribas, Raymond  \& Anglada-Escud{\'{e}}}{Turbet et~al.}{2016}]{Turbet2016}
Turbet M.,  Leconte J.,  Selsis F.,  Bolmont E.,  Forget F.,  Ribas I.,
  Raymond S.~N.,   Anglada-Escud{\'{e}} G.,  2016, \mn@doi [Astronomy {\&}
  Astrophysics] {10.1051/0004-6361/201629577}, 596, A112

\bibitem[\protect\citeauthoryear{Wertheimer \& Laughlin}{Wertheimer \&
  Laughlin}{2006}]{Wertheimer2006}
Wertheimer J.~G.,  Laughlin G.,  2006, \mn@doi [The Astronomical Journal]
  {10.1086/507771}, 132, 1995

\bibitem[\protect\citeauthoryear{Zubrin \& Andrews}{Zubrin \&
  Andrews}{1991}]{Zubrin1991}
Zubrin R.~M.,  Andrews D.~G.,  1991, \mn@doi [Journal of Spacecraft and
  Rockets] {10.2514/3.26230}, 28, 197

\makeatother
\end{thebibliography}



\appendix

\section{Analytic Solutions for Optimal Sail Orientation }\label{app:derivation}

\subsection{Sail Optimisation in 2D}

\noindent We wish to solve

\begin{equation}
\underset{\unitnormal \in\mathbb{R}^2 }{\text{minimize}} \,\,  \left(\unitnormal\cdot\mathbf{\hat{r}}\right)\left(\unitnormal\cdot\mathbf{\hat{v}}\right).
\end{equation}

\noindent Let 

\begin{equation}
\mathbf{\hat{r}} = \begin{pmatrix} 
\cos \phi \\ 
\sin \phi
\end{pmatrix}\, ,\, \mathbf{\hat{v}} = \begin{pmatrix} 
\cos \beta \\ 
\sin \beta \\
\end{pmatrix}
\end{equation}

\noindent The corresponding equation for the sail normal vector is

\begin{equation}
\mathbf{\hat{n}} = \begin{pmatrix} 
\cos (\alpha + \phi) \\ 
\sin (\alpha + \phi)
\end{pmatrix},
\end{equation}

\noindent where $\alpha$ is the angle between the sail normal and $\mathbf{\hat{r}}$:

\begin{equation}
\mathbf{\hat{r}}\cdot\mathbf{\hat{n}} = \cos \phi \cos (\alpha + \phi) + \sin \phi \sin(\alpha + \phi) = \cos \alpha.
\end{equation} 

\noindent Our minimisation condition is equivalent to minimising

\begin{equation}
f(\alpha) = \cos \alpha \left(\cos \beta \cos (\alpha + \phi) + \sin \beta \sin(\alpha + \phi)\right)
\end{equation}

\noindent with fixed $\phi$, $\beta$.  Appropriate use of double angle formulae gives

\begin{equation}
f(\alpha) = \cos \alpha \left( \cos (\alpha + \phi -\beta)\right).
\end{equation}

\noindent Taking the first derivative gives

\begin{multline}
f'(\alpha) = -\sin \alpha \left( \cos (\alpha + \phi -\beta)\right) - \cos \alpha \left( \sin (\alpha + \phi -\beta)\right) \\
 = - \sin \left(2\alpha + \phi -\beta\right)
\end{multline}

\noindent And the extrema of $f(\alpha)$ are to be found at

\begin{equation}
\alpha = \frac{\beta - \phi}{2}, \frac{\pi + \beta - \phi}{2}
\end{equation}

\noindent Taking the second derivative

\begin{equation}
f''(\alpha) = - 2\cos \left(2\alpha + \phi -\beta\right)
\end{equation}

\noindent shows that the minimum exists at

\begin{equation}
\alpha = \frac{\pi + \beta - \phi}{2}
\end{equation}

\subsection{Sail Optimisation in 3D}

\noindent To extend the above calculation to 3D, we must consider the vector field

\begin{equation}
\mathbf{f}(\mathbf{\hat{n}}) = \left(\unitnormal\cdot\mathbf{\hat{r}}\right)\left(\unitnormal\cdot\mathbf{\hat{v}}\right).
\end{equation}

\noindent And compute the gradient

\begin{equation}
\underline{\nabla} \mathbf{f} = \left(\frac{\partial \mathbf{f}}{\partial n_1}, \frac{\partial \mathbf{f}}{\partial n_2}, \frac{\partial \mathbf{f}}{\partial n_3}\right)
\end{equation}

\noindent where $n_i$ are the Cartesian components of the sail vector, and

\begin{equation}
\frac{\partial \mathbf{f}}{\partial n_i} = r_i \left(\unitnormal\cdot\mathbf{\hat{v}}\right) +v_i\left(\unitnormal\cdot\mathbf{\hat{r}}\right).
\end{equation}

\noindent Our minimum occurs where $\underline{\nabla} \mathbf{f}=0$.  This is equivalent to the homogeneous system of equations

\begin{equation}
T\mathbf{n} = 0
\end{equation}

\noindent where

\begin{equation}
T = \begin{pmatrix} 
2 r_1 v_1 & r_1 v_2 +r_2v_1 & r_1v_3 + r_3v_1 \\
r_2 v_1 +r_1 v_2 & 2r_2 v_2 & r_2 v_3 + r_3 v_2 \\
r_1 v_3 + r_3 v_1 & r_2 v_3 + r_3 v_2 & 2 r_3 v_3 \\
\end{pmatrix}
\end{equation}

\noindent Using this system in practice is challenging (especially in purely 2D trajectories), as $T$ can easily be singular and hence not be invertible.  To avoid these problems we use sequential least squares optimisation on this system as described in the main text.


\bsp	
\label{lastpage}
\end{document}